\documentclass[aps,pre,nofootinbib,showpacs,tightenlines,preprint,titlepage,amsmath,graphicx]{revtex4}
\usepackage{graphicx,color,epsfig,rotate}

\newcommand{\cC}{\ensuremath{\mathcal{C}}}
\newcommand{\cQ}{\ensuremath{\mathcal{Q}}}
\newcommand{\cP}{\ensuremath{\mathcal{P}}}
\newcommand{\cT}{\ensuremath{\mathcal{T}}}
\newcommand{\half}{\mbox{$\textstyle{\frac{1}{2}}$}}

\begin{document}
\title{Nonunique $\cC$ operator in $\cP\cT$ Quantum Mechanics}

\author{Carl M. Bender$^a$}
\email{cmb@wustl.edu}
\author{S. P. Klevansky$^b$}
\email{spk@physik.uni-heidelberg.de}
\affiliation{$^a$Physics Department, Washington University, St. Louis, MO 63130,
USA}
\affiliation{$^b$Institut f\"ur Theoretische Physik, Universita\"et Heidelberg,
Philosophenweg 19, 69120 Heidelberg, Germany}

\date{\today}

\begin{abstract} 
The three simultaneous algebraic equations, $\cC^2=1$, $[\cC,\cP\cT]=0$, $[\cC,H
]=0$, which determine the $\cC$ operator for a non-Hermitian $\cP\cT$-symmetric
Hamiltonian $H$, are shown to have a nonunique solution. Specifically, the $\cC$
operator for the Hamiltonian $H=\half p^2+\half\mu^2q^2+i\epsilon q^3$ is
determined perturbatively to first order in $\epsilon$ and it is demonstrated
that the $\cC$ operator contains an infinite number of arbitrary parameters. For
each different $\cC$ operator, the corresponding equivalent isospectral
Dirac-Hermitian Hamiltonian $h$ is calculated.
\end{abstract}

\maketitle
\section{Introduction}
\label{s1}

A Hamiltonian $H$ defines a physical theory of quantum mechanics if (i) $H$ has 
a real energy spectrum, and (ii) the time-evolution operator $U=e^{-iHt}$ is
unitary so that probability is conserved. These two features of the theory are
guaranteed if $H$ is Dirac Hermitian. (The Hamiltonian $H$ is {\em Dirac
Hermitian} if $H=H^\dag$, where the symbol $\dag$ represents the combined
operations of complex conjugation and matrix transposition.) However, it is not
necessary for $H$ to be Dirac Hermitian for the spectrum to be real and for time
evolution to be unitary. For example, the Hamiltonians belonging to the class
\begin{equation}
H=p^2+q^2(iq)^\epsilon\quad(\epsilon>0)
\label{e1}
\end{equation}
possess real eigenvalues \cite{S1,S2} and generate unitary time evolution
\cite{S3,S4}. These Hamiltonians therefore define physically acceptable quantum
theories.

The Hamiltonians in (\ref{e1}) are $\cP\cT$ symmetric, that is, invariant under
combined spatial reflection $\cP$ and time reversal $\cT$. The underlying reason
that the Hamiltonians $H$ in (\ref{e1}) are physically acceptable is that they
are selfadjoint, not with respect to the Dirac adjoint $\dag$, but rather with
respect to $\cC\cP\cT$ conjugation, where $\cC$ is a linear operator that
represents a hidden reflection symmetry of $H$. The $\cC\cP\cT$ inner product
defines a positive definite Hilbert space norm. Not every $\cP\cT$-symmetric
Hamiltonian has an entirely real spectrum but if the spectrum is entirely real,
a linear $\cP\cT$-symmetric operator $\cC$ exists that obeys the following three
algebraic equations \cite{S3}:
\begin{equation}
\cC^2=1,
\label{e2}
\end{equation}
\begin{equation}
\left[\cC,\cP\cT\right]=0,
\label{e3}
\end{equation}
\begin{equation}
\left[\cC,H\right]=0.
\label{e4}
\end{equation}
When such a $\cC$ operator exists, we say that the $\cP\cT$ symmetry of $H$ is
{\em unbroken}. Constructing the $\cC$ operator is the key step in showing that
time evolution for a non-Hermitian $\cP\cT$-symmetric Hamiltonian (\ref{e1}) is
unitary.

There has been much research activity during the past decade on non-Hermitian
$\cP\cT$-symmetric Hamiltonians \cite{S5} and the $\cC$ operator for some
nontrivial quantum-mechanical models has been calculated perturbatively
\cite{S6,S7,S8,S9}.
The first perturbative calculation of $\cC$ was performed in Ref.~\cite{S6} for
the Hamiltonian
\begin{equation}
H=\half p^2+\half\mu^2q^2+i\epsilon q^3.
\label{e5}
\end{equation}
Here, $\mu$ is a mass parameter, $\epsilon$ is a small coupling constant that is
used as a perturbation parameter, and $p$ and $q$ are the usual
quantum-mechanical operators satisfying the Heisenberg commutation relation
$[q,p]=i$. It was shown in Ref.~\cite{S6} that the $\cC$ operator has a simple
and natural form as the parity operator $\cP$ multiplied by an exponential of a
linear Dirac Hermitian operator $\cQ$:
\begin{equation}
\cC=e^\cQ\cP,
\label{e6}
\end{equation}
where $\cQ=\cQ^\dag$. For the Hamiltonian (\ref{e5}), $\cQ$ is a series in {\it
odd} powers of $\epsilon$:
\begin{equation}
\cQ=\cQ_1\epsilon+\cQ_3\epsilon^3+\cQ_5\epsilon^5+\cdots.
\label{e7}
\end{equation}
To first order in $\epsilon$ the result for $\cQ$ was found to be
\begin{equation}
\cQ_1=-\frac{2}{\mu^2}qpq-\frac{4}{3\mu^4}p^3.
\label{e8}
\end{equation}
Note that in the unperturbed limit $\epsilon\to0$ in which the Hamiltonian
becomes Hermitian and parity symmetry is restored, the operator $\cQ$ vanishes
and thus $\cC\to\cP$. This suggests that the $\cC$ operator may be interpreted
as the complex extension of the parity operator $\cP$.

Mostafazadeh showed that the $\cQ$ operator can be used to construct a
similarity transform that maps the non-Dirac-Hermitian Hamiltonian $H$ to a
spectrally equivalent Dirac-Hermitian Hamiltonian $h$ \cite{S10}:
\begin{equation}
h=e^{\cQ/2}He^{-\cQ/2}.
\label{e9}
\end{equation}
This similarity transformation was originally used by Scholtz {\it et al.} to
convert Hermitian Hamiltonians to non-Hermitian Hamiltonians \cite{S11}.
 
In Sec.~\ref{s2} of this paper we reinvestigate the Hamiltonian $H$ in
(\ref{e5}). We show that the perturbative solution in (\ref{e8}) for the $\cQ$
operator is in fact not unique and that the general solution to the algebraic
system (\ref{e2}) -- (\ref{e4}) contains an infinite number of arbitrary
continuous parameters. Then, in Sec.~\ref{s3}, we show that for each of these
$\cC$ operators there is a corresponding spectrally equivalent Hermitian
Hamiltonian $h$.

\section{Complete First-Order Calculation of the $\cC$ Operator}
\label{s2}

The purpose of this paper is to show that the $\cP\cT$-symmetric Hamiltonian $H$
in (\ref{e5}) has an infinite class of associated $\cC$ operators. To determine
the $\cC$ operator for a given Hamiltonian $H$ we must solve the three equations
(\ref{e2}) -- (\ref{e4}). The first two of these equations are kinematic
constraints on $\cC$; they are obeyed by the $\cC$ operator for any $\cP
\cT$-symmetric Hamiltonian. If we seek a solution for $\cC$ in the form
(\ref{e6}), we find that (\ref{e2}) and (\ref{e3}) imply that $\cQ(p,q)$ is an
{\it odd function of the} $p$ {\it operator and an even function of the $q$
operator.} The third equation (\ref{e4}) is dynamical because it makes explicit
reference to the Hamiltonian. Our specific objective here is to solve (\ref{e4})
perturbatively to first order in $\epsilon$.

We substitute (\ref{e5}) and (\ref{e6}) into (\ref{e4}) and to first order in
$\epsilon$ obtain the commutation relation satisfied by $\cQ_1$:
\begin{equation}
[\cQ_1,H_0]=2iq^3,
\label{e10}
\end{equation}
where $H_0=\half p^2+\half\mu^2q^2$ is the unperturbed Hamiltonian. The simplest
solution to (\ref{e10}) that is odd in $p$ and even in $q$ and is a {\it
polynomial} in $p$ and $q$ is given by the solution in (\ref{e8}). Note that
(\ref{e8}) is not the {\it unique} function of $p$ and $q$ that satisfies the
commutation relation (\ref{e10}) because we can add to $\cQ_1$ any function of
$H_0$. However, $H_0$ is an {\it even} function of $p$ and $\cQ_1$ is required
to be an {\it odd} function of $p$. Thus, one is tempted to conclude ({\it
wrongly}!) that the only solution to (\ref{e10}) is that given in (\ref{e8}).

To construct additional solutions to (\ref{e10}) we employ the operator notation
described in detail in Ref.~\cite{S12}. We introduce the set of totally
symmetric operators $T_{m,n}$, where $T_{m,n}$ is an equally weighted average
over all possible orderings of $m$ factors of $p$ and $n$ factors of $q$. For
example,
\begin{eqnarray}
T_{0,0}&=&1,\nonumber\\
T_{1,0}&=&p,\nonumber\\
T_{1,1}&=&\half(pq+qp),\nonumber\\
T_{1,2}&=&\textstyle{\frac{1}{3}}(pqq+qpq+qqp),\nonumber\\
T_{2,2}&=&\textstyle{\frac{1}{6}}(ppqq+qqpp+pqqp+qppq+qpqp+pqpq),
\label{e11}
\end{eqnarray}
and so on. Note that $T_{m,n}$ is the quantum-mechanical generalization of the
classical product $p^mq^n$. We can express the polynomial solution $\cQ_1$ in
(\ref{e8}) to the commutation relation (\ref{e10}) as a linear combination of
two such totally symmetric operators:
\begin{equation}
\cQ_1=-\frac{4}{3\mu^4}T_{3,0}-\frac{2}{\mu^2}T_{1,2}.
\label{e12}
\end{equation}

The advantage of the totally symmetric operators $T_{m,n}$ is that it is
especially easy to evaluate commutators and anticommutators. For example, as
shown in Ref.~\cite{S12}, the operators $T_{m,n}$ obey extremely simple
commutation and anticommutation relations:
\begin{eqnarray}
\left[p,T_{m,n}\right]&=&-inT_{m,n-1},\nonumber\\
\left[q,T_{m,n}\right]&=&imT_{m-1,n},\nonumber\\
\left\{p,T_{m,n}\right\}&=&2T_{m+1,n},\nonumber\\
\left\{q,T_{m,n}\right\}&=&2T_{m,n+1}.
\label{e13}
\end{eqnarray}
Also, by combining and iterating the results in (\ref{e13}), we can establish
additional useful commutation relations for $T_{m,n}$. For example,
\begin{eqnarray}
\left[p^2,T_{m,n}\right]&=&-2inT_{m,n-1},\nonumber\\
\left[q^2,T_{m,n}\right]&=&2imT_{m-1,n}.
\label{e14}
\end{eqnarray}

The totally symmetric operators $T_{m,n}$ can be re-expressed in Weyl-ordered
form \cite{S12}:
\begin{equation}
T_{m,n}=\frac{1}{2^m}\sum_{k=0}^m\binom{m}{k}p^kq^np^{m-k}=\frac{1}{2^n}
\sum_{k=0}^n\binom{n}{k}q^k p^m q^{n-k}\quad(m,\,n=0,\,1,\,2,\,3,\,\cdots).
\label{e15}
\end{equation}
The proof that the totally symmetric form of $T_{m,n}$ in (\ref{e11}) equals the
binomial-summation Weyl-ordered forms above requires the repeated use of the
Heisenberg algebraic property that $[q,p]=i$ and follows by induction. The
reason for introducing the Weyl-ordered form of $T_{m,n}$ is that it allows us
to extend the totally symmetric operators $T_{m,n}$ to negative values of $n$ by
using the first of these formulas or to negative values of $m$ by using the
second of these formulas. The commutation and anticommutation relations in
(\ref{e13}) and (\ref{e14}) remain valid if $m$ is negative or if $n$ is
negative.

We will now show that (\ref{e12}) is just one of many solutions to (\ref{e10})
that are even in $q$ and odd in $p$. Note that (\ref{e10}) is an {\it
inhomogeneous} linear equation for $\cQ_1$ and that (\ref{e12}) is a {\it
particular} solution to this equation. To find other solutions, we need only
solve the {\it associated homogeneous} linear equation
\begin{equation}
[\cQ_{1,\,{\rm homogeneous}},H_0]=0.
\label{e16}
\end{equation}
Let us look for a solution to this equation of the form
\begin{equation}
\cQ_{1,\,\rm homogeneous}=\sum_{k=0}^\infty a_k T_{3-2k,2k},
\label{e17}
\end{equation}
which by construction is odd in $p$ and even in $q$. Substituting (\ref{e17})
into (\ref{e16}) and using the commutation relations in (\ref{e14}), we obtain a
two-term recursion relation for the coefficients $a_n$:
\begin{equation}
a_{k+1}=-\frac{k-\frac{3}{2}}{k+1}\mu^2a_k\qquad(k=0,\,1,\,2,\,3,\,\cdots).
\label{e18}
\end{equation}
The solution to this recursion relation is
\begin{equation}
a_k=\frac{a}{\mu^4}\frac{\Gamma\left(k-\textstyle{\frac{3}{2}}\right)}{k!\,
\Gamma\left(-\textstyle{\frac{3}{2}}\right)}(-\mu^2)^k\qquad(k=0,\,1,\,2,\,3,\,
\cdots),
\label{e19}
\end{equation}
where $a$ is an arbitrary constant. Thus, a one-parameter family of solutions to
the homogeneous equation (\ref{e16}) is
\begin{equation}
\cQ_{1,\,\rm homogeneous}=\frac{a}{\mu^4}\sum_{k=0}^\infty\frac{\Gamma\left(k-
\frac{3}{2}\right)}{k!\,\Gamma\left(-\frac{3}{2}\right)}\left(-\mu^2\right)^k
T_{3-2k,2k}.
\label{e20}
\end{equation}
We have included the factor of $\mu^4$ in the denominator so that $\epsilon\cQ_{1,\,\rm
homogeneous}$ is dimensionless.

\begin{figure}[b!]
\vspace{3.7in}
\includegraphics{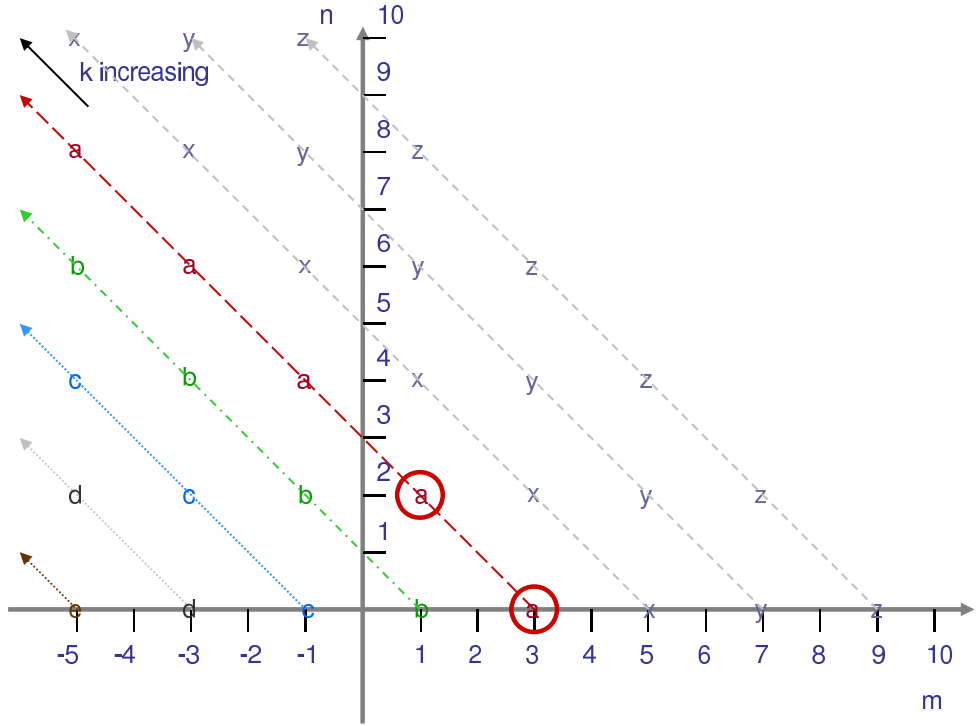}
\caption{Schematic representation of the coefficients in the operator series
solution for $Q_1$ in (\ref{e31}). The two circles represent the terms in the
inhomogeneous polynomial solution in (\ref{e12}). The $a$'s represent the
coefficients in the operator series (\ref{e20}), which is the homogeneous
solution corresponding to $P=0$. The coefficients of the homogeneous solutions
in (\ref{e30}) for $P=-1$, $-2$, and $-3$ are indicated by $b$'s, $c$'s, and
$d$'s, and the coefficients in the homogeneous solutions for $P=1$, $2$, and
$3$ are indicated by $x$'s, $y$'s, and $z$'s.}
\label{f1}
\end{figure}

Combining the inhomogeneous solution in (\ref{e12}) and the homogeneous
solutions in (\ref{e20}) gives a general one-parameter class of solutions to the
commutator condition (\ref{e10}):
\begin{equation}
\cQ_1=-\frac{4}{3\mu^4}T_{3,0}-\frac{2}{\mu^2}T_{1,2}+\frac{a}{\mu^4}\sum_{k=0
}^\infty\frac{\Gamma\left(k-\frac{3}{2}\right)}{k!\,\Gamma\left(-\frac{3}{2}
\right)}\left(-\mu^2\right)^kT_{3-2k,2k}.
\label{e21}
\end{equation}
The coefficients in this series are indicated schematically by the diagonal line
of a's in Fig.~\ref{f1} that intersects the $m$ axis at $(3,0)$ and the $n$
axis at $(0,3)$. On this figure we use circles to indicate the two terms in
(\ref{e21}) that arise from the particular polynomial solution (\ref{e12}).

One might be concerned that because the operator $p$ in the series in
(\ref{e21}) is raised to arbitrarily high negative powers it might be difficult
to interpret the series as an operator. However, we will now show how to sum
this operator series to obtain a well defined and meaningful result. First, we
observe that there is an extremely simple way to represent the totally symmetric
operator $T_{-n,n}$ for all integers $n$:
\begin{equation}
T_{-n,n}=\frac{1}{2}\left(q\frac{1}{p}\right)^n+\frac{1}{2}\left(\frac{1}{p}q
\right)^n.
\label{e22}
\end{equation}
This nontrivial formula can be verified by using the Heisenberg algebra $[q,p]=
i$. One may regard this formula as the companion to the result that
\begin{equation}
T_{n,n}={\rm Hahn}_n\left(T_{1,1}\right),
\label{e23}
\end{equation}
where ${\rm Hahn}_n(x)$ is a Hahn polynomial of degree $n$ \cite{S13}.

In order to use the formula (\ref{e21}), we replace the $k$th term in the series
in (\ref{e21}) with a triple anticommutator of $T_{-2k,2k}$:
\begin{equation}
T_{3-2k,2k}=\textstyle{\frac{1}{8}}\{\{\{T_{-2k,2k},p\},p\},p\}.
\label{e24}
\end{equation}
We then recognize that the sum is just a pair of binomial expansions of the
general form
\begin{equation}
(1+x)^\alpha=\sum_{k=0}^\infty\frac{\Gamma(k-\alpha)}{k!\,\Gamma(-\alpha)}
(-x)^k
\label{e25}
\end{equation}
with $\alpha=\frac{3}{2}$, one for each of the two operators $q\frac{1}{p}$ and
$\frac{1}{p}q$. Thus, the one-parameter family of solutions in (\ref{e21})
simplifies to
\begin{eqnarray}
\cQ_1&=&-\frac{4}{3\mu^4}T_{3,0}-\frac{2}{\mu^2}T_{1,2}+\frac{a}{16\mu^4}\left\{
\left\{\left\{\left(1+\mu^2q\frac{1}{p}q\frac{1}{p}\right)^{3/2},p\right\},p
\right\},p\right\}\nonumber\\
&&\qquad+\frac{a}{16\mu^4}\left\{\left\{\left\{\left(1+\mu^2\frac{1}{p}q\frac{1}
{p}q\right)^{3/2},p\right\},p\right\},p\right\}.
\label{e26}
\end{eqnarray}
Note that at the classical level, where $q$ and $p$ commute, the $p$'s in the
anticommutators combine with the expression to the power $\frac{3}{2}$ to give
$H_0^{3/2}$, which is even in $p$. However, in (\ref{e26}) we can see that the
symmetry requirement that $\cQ$ be odd in $p$ and even in $q$ is satisfied!
Furthermore, there are just enough powers of $p$ in the triple anticommutators
to cancel any small-$p$ singularities. 

One might now be tempted to argue as follows: In the Heisenberg algebra a
commutator behaves as a derivative. Thus, the commutator equation (\ref{e10}) is
analogous to a first-order linear ordinary differential equation. Since the
solution to such an equation contains one arbitrary parameter, the solution in
(\ref{e26}), which contains the arbitrary parameter $a$, should be the complete
solution to (\ref{e10}). However, this argument is wrong and we will now show
that there are in fact an {\it infinite number} of additional one-parameter
families of solutions to (\ref{e10}).

We seek new one-parameter families of solutions to the associated homogeneous
commutation relation (\ref{e16}), with each family of solutions labeled by the
integer $P=0,\,\pm1,\,\pm2,\,\pm3,\,\cdots$. For each value of $P$ the solution
has the form
\begin{equation}
\cQ_{1,\,\rm homogeneous}^{(P)}=\sum_{k=0}^\infty a_k^{(P)}T_{2P+3-2k,2k},
\label{e27}
\end{equation}
which is odd in $p$ and even in $q$, as is required. We substitute (\ref{e27})
into (\ref{e16}) and use the commutation relations (\ref{e14}) to obtain a
recursion relation for the coefficients $a_k^{(P)}$:
\begin{equation}
a_{k+1}^{(P)}=-\frac{k-P-\frac{3}{2}}{k+1}\mu^2a_k^{(P)}\qquad(k=0,\,1,\,2,\,3,
\,\cdots).
\label{e28}
\end{equation}
The solution to this recursion relation is
\begin{equation}
a_k^{(P)}=\frac{a^{(P)}}{\mu^{P+4}}\frac{\Gamma\left(k-P-\textstyle{\frac{3}{2}}
\right)}{k!\,\Gamma\left(-P-\textstyle{\frac{3}{2}}\right)}(-\mu^2)^k\qquad
(k=0,\,1,\,2,\,3,\,\cdots),
\label{e29}
\end{equation}
where $a^{(P)}$ are arbitrary constants. Thus, for each integer $P$ we obtain
the following one-parameter family of solutions to the homogeneous equation
(\ref{e16}):
\begin{equation}
\cQ_{1,\,\rm homogeneous}^{(P)}=\frac{a^{(P)}}{\mu^{P+4}}\sum_{k=0}^\infty\frac{
\Gamma\left(k-P-\frac{3}{2}\right)}{k!\,\Gamma\left(-P-\frac{3}{2}\right)}\left
(-\mu^2\right)^k T_{2P+3-2k,2k}.
\label{e30}
\end{equation}
The factor of $\mu^{P+4}$ in the denominator ensures that $\epsilon\cQ_{1,\,\rm
homogeneous}^{(P)}$ is dimensionless. Note that if we set $P=0$, we recover the
homogeneous solution in (\ref{e20}). In Fig.~\ref{f1} the coefficients for the
$P=-1$, $P=-2$, and $P=-3$ solutions are indicated by $b$'s, $c$'s, and $d$'s
and the coefficients for the $P=1$, $P=2$, and $P=3$ solutions are indicated by
$x$'s, $y$'s, and $z$'s.

We can now combine all these solutions to produce a very general solution
for $Q_1^{(P)}$:
\begin{equation}
\cQ_1=-\frac{4}{3\mu^4}T_{3,0}-\frac{2}{\mu^2}T_{1,2}+\sum_{P=-\infty}^\infty
\frac{a^{(P)}}{\mu^{P+4}}\sum_{k=0}^\infty\frac{\Gamma\left(k-P-\frac{3}{2}
\right)}{k!\,\Gamma\left(-P-\frac{3}{2}\right)}\left(-\mu^2\right)^kT_{2P+3-2k,
2k}.
\label{e31}
\end{equation}
This solution generalizes and replaces that in (\ref{e21}) and is one of the
principal results in this paper. We emphasize that $a^{(P)}$ are all arbitrary.

Finally, we observe that it is possible to present the infinite sums over
operators in (\ref{e30}) and (\ref{e31}) more compactly by performing the sum
over $k$. To do so, we generalize the result in (\ref{e24}) as a $2P+3$-fold
anticommutator for the case when $P\geq-1$,
\begin{equation}
T_{2P+3-2k,2k}=\textstyle{\frac{1}{2^{2P+3}}}\{\{\ldots\{T_{-2k,2k},p\},\ldots
p\},p\}_{2P+3\,{\rm times}}.
\label{e32}
\end{equation}
This result allows us to apply the identity in (\ref{e25}) to perform the sum in
(\ref{e30}). We find that
\begin{eqnarray}
\cQ_{1,\,\rm homogeneous}^{(P)}&=&\frac{a^{(P)}}{2^{2P+4}\mu^{P+4}}\left\{\left
\{\ldots\left\{\left(1+\mu^2q\frac{1}{p}q\frac{1}{p}\right)^{P+3/2},p\right\},
\ldots p\right\},p\right\}_{2P+3\,{\rm times}}\nonumber\\
&&\!\!\!\!\!\!\!\!\!\!\!\!\!\!\!+\frac{a^{(P)}}{2^{2P+4}\mu^{P+4}}\left\{\left\{
\ldots\left\{\left(1+\mu^2\frac{1}{p}q\frac{1}{p}q\right)^{P+3/2},p\right\},
\ldots p\right\},p\right\}_{2P+3\,{\rm times}}.
\label{e33}
\end{eqnarray}
One may simplify this expression even further by exploiting the analog of the
Baker-Hausdorff formula for multiple anticommutators \cite{S14}:
\begin{equation}
e^ABe^A=B+\{A,B\}+\frac{1}{2!}\{A,\{A,B\}\}+\frac{1}{3!}\{A,\{A,\{A,B\}\}\}
+\cdots.
\label{e34}
\end{equation}
Thus, the $2P+3$-fold anticommutator in (\ref{e32}), for example, may be
rewritten in a more compact form as a multiple derivative:
\begin{equation}
\{\{\ldots\{T_{-2k,2k},p\},\ldots p\},p\}_{2P+3\,{\rm times}}=\frac{d^{2P+3}}
{d\beta^{2P+3}}e^{\beta p}\,T_{-2k,2k}\,e^{\beta p}\bigg|_{\beta=0}.
\label{e35}
\end{equation}
When $P<-1$, we can use commutators instead of anticommutators to make the
expressions more compact, but we do not present the results here because
they are repetitive.

\section{Spectrally Equivalent Hamiltonians}
\label{s3}

For the $\cC$ operators associated with $Q_1$ in (\ref{e31}) we must now
calculate the Hermitian Hamiltonian $h$ that is equivalent (in the sense that it
is isospectral) to the Hamiltonian in $H$ in (\ref{e5}). To do so, we evaluate
the similarity transformation in (\ref{e9}), which to leading order in
$\epsilon$ amounts to evaluating the commutator
\begin{equation}
h=H_0+\textstyle{\frac{1}{4}}i\epsilon^2\left[q^3,\cQ_1\right].
\label{e36}
\end{equation}
Note that when we use the {\it first-order} form of the $\cQ$ operator, we
obtain $h$ to {\it second order} in $\epsilon$. (The third-order contribution
to $\cQ$ gives $h$ to fourth order in $\epsilon$.)

In order to evaluate the commutator in (\ref{e36}) we use the relation 
\begin{equation}
\left[q^3,T_{m,n}\right]=-\textstyle{\frac{1}{4}}im(m-1)(m-2)T_{m-3,n}
+3imT_{m-1,n+2},
\label{e37}
\end{equation}
which is derived from the commutation relations in (\ref{e13}) and (\ref{e14}).
Furthermore, since (\ref{e36}) depends on $Q_1$ linearly, we can evaluate each
of the contributions to $Q_1$ independently and add together the results at
the end of the calculation.

First, we evaluate the commutator in (\ref{e36}) for the inhomogeneous part (the
first two terms) in $Q_1$ and obtain
\begin{equation}
h=H_0+\frac{\epsilon^2}{2\mu^4}\left(-1+6T_{2,2}+3\mu^2T_{0,4}\right).
\label{e38}
\end{equation}
This result was already obtained in Ref.~\cite{S15}, where it was observed that
to this order in perturbation theory the equivalent Hermitian Hamiltonian $h$
represents a quartic anharmonic oscillator having a position-dependent mass.
However, this interpretation holds only for the case in which we take the
coefficients $a^{(P)}$ to vanish for all $P$.

For each nonzero value of $a^{(P)}$ the contribution to the commutator in
(\ref{e36}) for $\cQ_{1,\,\rm homogeneous}^{(P)}$ in (\ref{e30}) is
\begin{eqnarray}
\frac{1}{4}i\epsilon^2\left[q^3,\cQ_{1,\,\rm homogeneous}^{(P)}\right]&=&
\frac{i\epsilon^2a^{(P)}}{4\mu^{P+4}}\sum_{k=0}^\infty\frac{\Gamma\left(k-P-
\frac{3}{2}\right)}{k!\,\Gamma\left(-P-\frac{3}{2}\right)}\left(-\mu^2\right)^k
\left[q^3,T_{2P+3-2k,2k}\right]\nonumber\\
&=&\frac{\epsilon^2a^{(P)}}{2\mu^{P+4}}\sum_{k=0}^\infty\frac{(-\mu^2)^k}{k!}
\nonumber\\
&& \!\!\!\!\!\!\!\!\!\!\!\!\!\!\!\!\!\!\!\!\!\!\!\!\!\!\!\!\!\!\!\!\!\!\!\!\!\!
\!\!\!\!\!\!\!\!\!\!\!\!\!\!\!\!\!\!\!\! \times\left[\frac{(P+1-k)\Gamma\left(
k-P+\frac{1}{2}\right)}{\Gamma\left(-P-\frac{3}{2}\right)}T_{2P-2k,2k}-\frac{3k
\Gamma\left(k-P-\frac{3}{2}\right)}{\mu^2\Gamma\left(-P-\frac{3}{2}\right)}
T_{2P+4-2k,2k}\right],
\label{e39}
\end{eqnarray}
where we have made use of (\ref{e37}).

We have shown in this paper that the $\cC$ operator contains an infinite number
of arbitrary parameters. It remains an open question as to whether there is an
additional physical or mathematical condition that would determine the $\cC$
operator uniquely; that is, whether there is an advantageous or a ``best''
choice for the arbitrary parameters $a^{(P)}$ in (\ref{e31}). For example, one
might try to choose $a^{(P)}$ such that the equivalent Hermitian Hamiltonian $h$
has the form $p^2+V(q)$. [The non-Dirac-Hermitian $\cP\cT$-symmetric Hamiltonian
$H=p^2-gq^4$ is spectrally equivalent to the Dirac-Hermitian Hamiltonian $h=p^2+
4gq^4-2\sqrt{g}\hbar q$ \cite{S16,S17,S18,S19,S20}.] However, there appears to
be no choice of the parameters $a^{(P)}$ that achieves such a simple form. Thus,
we conclude by posing the question, is there a useful fourth constraint that one
can use to supplement the three conditions in (\ref{e2}), (\ref{e3}), and
(\ref{e4}), that gives rise to the ``best form'' for the $\cC$ operator?

We conjecture that the answer to this question in the context of quantum field
theory is that the huge parametric freedom in the $\cC$ operator can be used to
impose the condition of {\it locality}. A $\cC$ operator for a $g\varphi^3$
quantum field theory has been calculated perturbatively to first order in $g$
\cite{S4}. The resulting $\cQ$ operator was found to decay exponentially at
large spatial distances because it contains an associated Bessel function. It
may be possible to exploit the parametric freedom in $\cC$ to replace the Bessel
function by a spatial delta function so that the $\cC$ operator becomes local.

\acknowledgments{CMB is grateful to the Graduate School of the University of
Heidelberg, where this work was done, for its hospitality. CMB thanks the
U.S.~Department of Energy for financial support.}


\begin{thebibliography}{99}

\bibitem{S1} C. M. Bender and S. Boettcher, Phys. Rev. Lett. {\bf 80},
5243 (1998).

\bibitem{S2} P.~Dorey, C.~Dunning and R.~Tateo, J.~Phys.~A {\bf 34} L391
(2001); {\em ibid}. {\bf 34}, 5679 (2001).

\bibitem{S3} C.~M.~Bender, D.~C.~Brody, and H.~F.~Jones, Phys.~Rev.~Lett.~{\bf
89}, 270401 (2002) and Am.~J.~Phys.~{\bf 71}, 1095 (2003).

\bibitem{S4} C.~M.~Bender, D.~C.~Brody, and H.~F.~Jones, Phys.~Rev.~Lett.~{\bf
93}, 251601 (2004) and Phys.~Rev.~D {\bf 70}, 025001 (2004).

\bibitem{S5} See, for example, G.~Levai and M.~Znojil, J.~Phys.~A:
Math.~Gen.~{\bf 33}, 7165 (2000).

\bibitem{S6} C.~M.~Bender, P.~N.~Meisinger, and Q.~Wang, J.~Phys.~A:
Math.~Gen.~{\bf 36}, 1973 (2003).

\bibitem{S7} C.~M.~Bender and B.~Tan, J.~Phys.~A: Math.~Gen.~{\bf 39}, 1945
(2006).

\bibitem{S8} K.~A.~Milton, Czech.~J.~Phys.~{\bf 54}, 85 (2004); C.~M.~Bender,
I.~Cavero-Pelaez, K.~A.~Milton, and K.~V.~Shajesh, Phys.~Lett.~B {\bf 613}, 97
(2005).

\bibitem{S9} C.~M.~Bender and H.~F.~Jones, Phys.~Lett.~A {\bf 328}, 102 (2004);
C.~M.~Bender, J.~Brod, A.~Refig, and M.~E.~Reuter, J.~Phys.~A: Math.~Gen.~{\bf
37}, 10139 (2004);
C.~M.~Bender, S.~F.~Brandt, J.-H.~Chen, and Q.~Wang, Phys.~Rev.~D {\bf 71},
025014 (2005);
C.~M.~Bender, H.~F.~Jones and R.~J.~Rivers, Phys.~Lett.~B {\bf 625}, 333 (2005);
C.~M.~Bender, S.~F.~Brandt, J.-H.~Chen, and Q.~Wang, Phys.~Rev.~D {\bf 71},
065010 (2005).

\bibitem{S10} A.~Mostafazadeh, J.~Math.~Phys.~{\bf 43}, 205 (2002); J.~Phys.~A:
Math.~Gen.~{\bf 36}, 7081 (2003).

\bibitem{S11} F.~Scholtz, H.~Geyer, and F.~Hahne, Ann.~Phys.~{\bf 213}, 74
(1992).

\bibitem{S12} C.~M.~Bender and G.~V.~Dunne, Phys.~Rev.~D {\bf 40}, 10 (1989).

\bibitem{S13} C.~M.~Bender, L.~R.~Mead, and S.~S.~Pinsky, J.~Math.~Phys.~{\bf
28}, 509 (1987).

\bibitem{S14} I.~Menda\v{s} and P. Milutinovi\'c, J.~Phys.~A: Math.~Gen.~{\bf
22}, L687 (1989).

\bibitem{S15} A.~Mostafazadeh, J.~Phys.~A: Math.~Gen.~{\bf 38}, 6557 (2005);
Erratum, {\it ibid.} {\bf 38}, 8185 (2005).

\bibitem{S16} A.~A.~Andrianov, Ann.~Phys.~{\bf 140}, 82 (1982).

\bibitem{S17} V.~Buslaev and V.~Grecchi, J.~Phys.~A: Math.~Gen.~{\bf 26}, 5541
(1993).  

\bibitem{S18} H.~F.~Jones and J.~Mateo, Phys.~Rev.~D {\bf 73}, 085002 (2006).

\bibitem{S19} C.~M.~Bender, D.~C.~Brody, J.-H.~Chen, H.~F. Jones, K.~A.~Milton,
and M.~C.~Ogilvie, Phys. Rev. D {\bf 74}, 025016 (2006).

\bibitem{S20} A.~A.~Andrianov, Phys.~Rev.~D {\bf 76}, 025003 (2007).

\end{thebibliography}
\end{document}